\input harvmac

\Title{\vbox{\baselineskip12pt
\hbox{BCCUNY-HEP/02-02} \hbox{hep-th/0202158}}}
{\vbox{\centerline{Strings, Fivebranes and an Expanding Universe}}}
\baselineskip=12pt
\centerline {Ramzi R. Khuri$^{1,2,3}$\footnote{$^a$}{e-mail: khuri@gursey.baruch.cuny.edu.}
and Andriy Pokotilov$^{1,2}$\footnote{$^b$}{e-mail: APokotilov@gc.cuny.edu.
}}
\medskip
\centerline{\sl $^1$Department of Natural Sciences, Baruch College,
CUNY} \centerline{\sl 17 Lexington Avenue, New York, NY 10010}
\medskip
\centerline{\sl $^2$Graduate School and University Center, CUNY}
\centerline{\sl 365 5th Avenue, New York, NY 10036}
\medskip
\centerline{\sl $^3$Center for Advanced Mathematical Sciences}
\centerline{\sl American University of Beirut, Beirut, Lebanon
\footnote{$^{**}$}{Associate member.}}

\bigskip
\centerline{\bf Abstract}
\medskip
\baselineskip = 20pt

It was recently shown that velocity-dependent forces between parallel fundamental
strings moving apart in a $D-$dimensional spacetime implied an accelerating expanding universe
in  $D-1$-dimensional space-time. Exact solutions were obtained for the early time expansion 
in $D=5,6$. Here we show that this result also holds for fundamental strings in the 
background of a fivebrane, and argue that the feature of an accelerating universe would
hold for more general $p$-brane-seeded models. 

\Date{February 2002}

\def\dx{\dot x}

\def\a{\alpha}

\def\r{\rho}

\def\({\left (}
\def\){\right )}
\def\[{\left [}
\def\]{\right ]}

\lref\chams{A. H. Chamseddine, Phys. Rev. {\bf D24} (1981) 3065.}

\lref\cvet{M. Cvetic and D. Youm, Phys. Rev. {\bf D53 } (1996) 584 ;
M. Cvetic and A. A. Tseytlin, Phys. Rev. {\bf D53} (1996) 5619.}

\lref\comp{R. R. Khuri, hep-th/9609094.}

\lref\prep{M. J. Duff, R. R. Khuri and J. X. Lu, Phys. Rep.
{\bf B259} (1995) 213, hep-th/9412184.}

\lref\dab{A. Dabholkar, G. W. Gibbons, J. A. Harvey and F. Ruiz Ruiz,
Nucl. Phys. {\bf B340} (1990) 33.}

\lref\fundst{R. R. Khuri, Phys. Lett. {\bf B353} (2001) 520, hep-th/0109041.}

\lref\witten{E. Witten, hep-th/0106109 and references therein.}

\lref\duffl{M. J. Duff and J. X. Lu, Nucl. Phys. {\bf B354} (1991) 141.}

\lref\dufflu{M. J. Duff and J. X. Lu, Nucl. Phys. {\bf B354} (1991) 129.}

\lref\our{R. R. Khuri and A. Pokotilov, hep-th/0201194}

\lref\dkl{M. J. Duff, R. R. Khuri and J. X. Lu, Nucl. Phys. {\bf B377} (1992) 281.}

\lref\calk{C. G. Callan and R. R. Khuri, Phys. Lett. {\bf B261} (1991) 263.}


In recent work \refs{\fundst,\our}, it was shown that velocity-dependent forces
between parallel fundamental strings moving apart in a $D$-dimensional spacetime implied
an accelerating expanding universe in $D-1$-dimensional spacetime. Exact
solutions were obtained for the expansion rate for simple models in $D=5,6$ \our.  
It was also noted in \our\ that an accelerating universe can equally well arise
for parallel $p$-branes, for arbitrary $p$, preserving half the spacetime supersymmetries
and moving apart in the transverse space.

In this letter we investigate whether the feature of an accelerating universe holds
for different types of $p$-branes. We first verify that the string result holds for
parallel fivebranes moving apart and then consider strings moving in a fivebrane
background. We find an explicit solution in the mean-field approximation for an 
expanding, accelerating universe consisting of strings moving in the fivebrane background.

For the fundamental string solution \dab, we start with the combined action 
$S_{string}=I_D + S_2$, where 
\eqn\sgact{I_D = {1\over 2\kappa_D^2} \int d^D x \sqrt{-g}
 \left(R- {1 \over 2}(\partial\phi)^2 -{1\over 2 \cdot 3!} e^{-\phi} H_3^2\right)}
is the $D$-dimensional string low-energy effective spacetime action and
\eqn\smact{S_2 =-{T_2\over 2} \int
d^2 \zeta \left(\sqrt{-\gamma} \gamma^{\mu\nu} \partial_\mu X^M
\partial_\nu X^N g_{MN} e^{\phi/2} +
\epsilon^{\mu\nu} \partial_\mu X^M \partial_\nu X^N B_{MN}\right)}
is the two-dimesional worldsheet
sigma-model source action. Upper case latin leters
denote indices of target space-time, lower case greek leters
are worldvolume indices. $g_{MN}$, $B_{MN}$ and $\phi$ are the
spacetime canonical metric, antisymmetric tensor and dilaton,
respectively, while $\gamma_{\mu\nu}$ is the worldsheet metric.
$H_3 = dB_2$ and $T_2$ is the string tension equal to its mass/length.                                      
The string sigma-model action is related to the canonical metric via
$g^s_{MN} = e^{\phi/2} g_{MN}$. 

The fundamental string solution represents stationary macroscopic string
parallel to the $x^1$ direction and is given by \dab
\eqn\fstring{\eqalign{ds^2 & = e^{3 \phi/2} \left(-dt^2+(dx^1)^2\right)
+ e^{-\phi/2}\delta_{ij} dx^i dx^j,\cr
e^{-2\phi} & \equiv h = 1 + {k_n \over r^n},\qquad B_{01} = - h^{-1} \cr}}
where $n=D-4$ (we assume $D>4$), $r^2 = x^i x_i$ and
the lower case latin indices run through the $D-2$-dimensional
space transverse to the worldvolume ($i,j=2,3,\dots, D-1$). The constant 
$k_n = {2\kappa_D^2 T_2 \over n\Omega_{n+1}}$,
where $\Omega_{n+1}$ is the volume of the $n+1$-dimensional unit sphere.

The Lagrangian for a test fundamental string moving in
the background of a parallel source string is then given by \refs{\fundst,\calk}
\eqn\lag{{\cal L}_2=-m h^{-1} \left(\sqrt{1-h \dx^2} - 1\right),}
where $m$ is the mass of the string, $\dx ^2 = \dx^i \dx_i$ and
the ``$\cdot$'' represents a time derivative. This Lagrangian does not 
depend explicitly on time, so that the total energy of the system is 
conserved. For $D=5,6$, the equation for conservation of energy can be
integrated exactly \our. 

For $D=5$ a straightforward integration yields 
\eqn\meanone{\left({\r +3 \over \r +1}\right)\ln \left( \sqrt{r\over a} + \sqrt{{r\over a}+1}\right)
+
\sqrt{r\over a} \sqrt{{r\over a}+1}
= \sqrt{\left(\r+2\right)^3\over \r\left(\r+1\right)^2}
\left({t-t_0\over k_1}\right),}
where $\rho={E \over m}$, $a={\r k_1\over \r +2}$ and $t_0$ is a constant.
For small $r$ (or early time $t$) \our, 

\eqn\rsmall{r \simeq \left({\r +2 \over \r +1} \right)^2 {t^2 \over k_{1}},}
while for large $r$ (or late time), $r \propto t$.

For $D=6$, we obtain
\eqn\meantwo{ \sqrt{r^2 + a} +
\sqrt{a} \left( {\r+2\over \r +1} \right)
\ln \left( {r + \sqrt{r^2 +a} - \sqrt{a} \over r + \sqrt{r^2 +a} + \sqrt{a}}\right)
= \sqrt{{\r(\r +2)\over(\r +1)^2}} (t-t_0),}
where again $a={\r k_2\over \r +2}$ and $t_0$ is a constant.
For small $r$, 
\eqn\smr{r \simeq r_0 \exp{{t \over \sqrt{k_2}}},} 
while for large $r$ we again find
$r \propto t$. This late time behaviour is a general property 
of this kind of model and is valid for any $D$ \fundst.

In \our\ it was shown that some simple models of a string-seeded universe
in $D=5,6$ have the same early time behaviour \rsmall,\smr\ as corresponding 
source/test string systems. This suggests that the mean-field approximation
of \fundst\ provides a valid description of the early time expansion rate for these systems.
Late time expansion rates for these models can also be 
obtained from conservation of energy and may represent a testable 
prediction (see \our\ for further discussions).

The fundamental string in $D=10$ is a solution of $3-$form version of
$D=10, N=1$ supergravity. The dual $7-$form version of this theory \chams\
corresponds to supergravity coupled to the fivebrane $\sigma-$model 
\refs{\duffl,\prep}. The action for the supergravity fields $(g_{MN},
A_{MNPQRS}, \phi)$ is now

\eqn\five{I_{10} = {1\over 2\kappa_{10}^2} \int d^{10} x \sqrt{-g}
 \left(R- {1 \over 2}(\partial\phi)^2 -{1\over 2 \cdot 7!} e^{\phi} K_7^2\right)}
where $K_7=dA_6$. $I_{10}$ is the same action as $I_D$ in \sgact\ for $D=10$
provided $H_3$ and $K_7$ are related via the duality transformation
\eqn\duality{K_7 = {} ^* H_3 {} e^{-\phi}.} 
The fivebrane sigma-model action is given by \refs{\duffl, \dufflu}

\eqn\sigma{\eqalign{S_6 =-T_6 \int
d^6 \zeta \left(\sqrt{-\gamma} \gamma^{\mu\nu} \partial_\mu X^M
\partial_\nu X^N g_{MN} e^{-\phi/6} - 2 \sqrt{-\gamma}+ \right. \cr
\left.
{1 \over 6!}\epsilon^{\mu_1 \mu_2 \dots
\mu_6} \partial_{\mu_1} X^{M_1}
\partial_{\mu_2} X^{M_2} \partial_{\mu_3} X^{M_3} 
\partial_{\mu_4} X^{M_4} \partial_{\mu_5} X^{M_5}
\partial_{\mu_6} X^{M_6} 
A_{M_1 M_2\dots 
M_6}\right)\cr}}
where $T_6$ is the fivebrane tension. The fivebrane sigma model metric is related
to the canonical metric via $g^f_{MN} = e^{-\phi/6} g_{MN}$.

The fundamental fivebrane solution to the equations of motion of the combined action 
$S_{fivebrane}=I_{10}+S_6$ is given by
\eqn\ffive{\eqalign{ds^2 & = e^{-\phi/2} \left(-dt^2+(dx^1)^2\right)
+ e^{3 \phi/2}\delta_{mn} dx^m dx^n,\cr
e^{2\phi} & = 1 + {\tilde k_2 \over r^2},\qquad A_{012345} = - e^{-2\phi} \cr}}
where $\tilde k_2 = {\kappa_{10}^2 T_6\over \Omega_3}$,
$m,n=6,7,8,9$ and $r$ is the radial coordinate in the four-dimensional space 
transverse to the six-dimensional worldvolume of the fivebrane.

The Lagrangian for a test fivebrane moving in the background of a parallel 
source-fivebrane \dkl\ can be easily obtained from \sigma, \ffive\ and
in term of the proper time of the test fivebrane takes the form
\eqn\fivefive{{\cal L}_6=-m e^{-2\phi} \left(\sqrt{1-e^{2 \phi} \dot{r}^2} - 1\right),}
where $m$ is the mass of the test fivebrane.
Thus the Lagrangian \fivefive\ for this two-fivebrane system in $D=10$ has exactly
the same form as the Lagrangian for the test string moving in the source string
background in $D=6$. This fact is a consequence of string/fivebrane
duality in $D=10$ or even string/string duality in $D=6$, once the fivebrane
is reduced to a dual string in $D=6$ \prep. Note that we could have obtained the identical
result as above starting directly with the dual string in $D=6$.
So the early time dependence $r=r(t)$ (where r is the 
radial coordinate for the four-dimensional space transverse to the six-dimensional
worldvolume) has the form $r \simeq r_0 \exp{{t \over \sqrt{\tilde k_2}}}$ where $r_0$ is the 
initial distance between the fivebranes in this four-dimensional space.
In the mean-field approximation, the position of the test-fivebrane represents the
average fivebrane-fivebrane distance and hence the scale size of the universe (see
\refs{\fundst,\our}).

We now wish to investigate whether the above scenario holds for velocity-dependent
forces between different branes. The most natural case to consider is that of a brane
propagating in the background of a dual brane. In particular, 
we consider a test string moving in the background of a fivebrane. Suppose
the fivebrane is oriented along $x^{\a}=\zeta^{\a}$ ($\a=1,2,\dots,5$). We assume
that the test string lies either parallel or antiparallel to one of the fivebrane
directions, say $x^1$. Viewed as a background for string propagation, the fivebrane
is a nonsingular solution of the spacetime action $I_{10}$ alone, without the need
for a source term (since no singularity is present in the string frame).
The metric and dilaton are then given by \ffive, with the three-form
given by the duality transformation \duality. Since, from
\ffive, the only nonvanishing components of $K$ are of the form $K_{012345m}$
where the directions $m=6,7,8,9$ are transverse to the fivebrane, by dualizing
we obtain that the only nonzero components of $H_3=dB_2$ are $H_{pqs}(r)$
where again $p,q,s=6,7,8,9$. Thus the only nonvanishing components of $B_{MN}$
occur when $M,N=6,7,8,9$. It then follows that the Wess-Zumino-Witten term
in the action \smact\ for the test string in the fivebrane background 
$\epsilon^{\mu\nu} \partial_\mu X^M \partial_\nu X^N B_{MN}$
vanishes. Replacing the fields of the fivebrane from
\ffive\ into \smact\ we obtain the Lagrangian \dkl 
\eqn\strfive{{\cal L}=-m \sqrt{1- e^{2\phi} \dot{r}^2}}
where $m$ is the mass of the test string and $e^{2 \phi}$ is given by \ffive.
The Hamiltonian of this test string/source fivebrane system also does not depend on 
time and thus represents the conserved energy
\eqn\hamfive{E \equiv H={m \over \sqrt{1-e^{2\phi} \dot{r}^2}}.}
Integrating \hamfive\ over time we obtain
\eqn\consener{\sqrt{r^2+\tilde k_2}-\sqrt{\tilde k_2} 
\ln{\sqrt{r^2+\tilde k_2} +\sqrt{\tilde k_2} \over r}=\sqrt{\a}t+const,}
where $\a=1-{m^2 \over E^2}$.

For early times $(r<< \sqrt{\tilde k_2})$ we obtain from \consener
\eqn\smallrfive{r \simeq r_0 \exp{\left( \sqrt{\a \over \tilde k_2}t \right)}}
i.e. the same type of early time dependence as for parallel strings in 
$D=6$. Once again, for late times we obtain $r\propto t$ 
as a general feature of this type of model \refs{\fundst, \our}.
We would expect this feature to persist for any dual pair of $p$-brane, in 
particular for fivebranes propagating in the background of a string.

Since similar results seem to hold for branes propagating in either identical
or dual brane backgrounds, this strongly suggests that the type of velocity-dependent forces
present in the above cases are generic to branes and would lead to an accelerating expanding universe
for various different types of branes simultaneously propagating in the same background.
This is also supported by the compositeness feature of brane solutions, which allows for
the construction of arbitrary brane solution from fundamental brane building blocks (of which
the fundamental string and fivebrane are examples) \refs{\comp,\cvet}. Another interesting question
is whether these scenarios hold in the context of general relativity, independent of string
theory. A possible extension of these results is to consider different
orientations of the various branes, in which case the zero-force condition no longer holds in the static
limit. Finally, it is worthwhile to go beyond the mean-field approximation and
investigate the many-body problem directly (see \our\ for some simplified models), 
also taking into account quantum interactions.

{\bf Acknowledgements:} Research supported by PSC-CUNY Grant \# 63497 00 32 and 
a Eugene Lang Junior Faculty Research Fellowship.

\listrefs
\end